\documentclass[a4 paper, 11pt,twoside]{article}
\usepackage{fullpage}                           
\usepackage[lmargin=0.6in,rmargin=0.6in,tmargin=0.6in,headsep=.2in]{geometry}
\usepackage{graphicx}
\usepackage{forloop}
\usepackage{etoolbox}
\usepackage{calc}
\usepackage{wrapfig,lipsum,booktabs} 
\usepackage{xtab} 
\usepackage[dvipsnames]{xcolor}
\usepackage[normalem]{ulem}
\usepackage{sectsty}
\allsectionsfont{\color{Brown}}
\makeatletter
\def\@seccntformat#1{\@ifundefined{#1@cntformat}%
{\csname the#1\endcsname\;}
{\csname #1@cntformat\endcsname}
}
\def\section@cntformat{\thesection.\;} 
\def\subsection@cntformat{\thesubsection.\;} 
\makeatother
\usepackage{hyperref}
\hypersetup{
    colorlinks=true,
    urlcolor=blue,
    linkcolor=black,
    citecolor = black}

\usepackage{fancyhdr}
\usepackage{amsmath,amsfonts,amssymb,amsthm,epsfig,epstopdf,url,array}
 \usepackage[retainorgcmds]{IEEEtrantools}
 \usepackage{makeidx,epsfig,lscape}
 \usepackage{xcolor,pict2e}
\usepackage{upgreek}

\begin{document}
\vspace*{3cm}
{\noindent\huge\bf A Simulation of the Dependence of Tidal Interaction on Galaxy Type in Compact Groups}\\
\\
{\bf\large Mark J. Henriksen, Mateo Mejia}\\[0.5cm]
University of Maryland, Baltimore County\\
Email: henrikse@umbc.edu\\
\\
{\color{Brown}\rule{0.7\textwidth}{2pt}}\\[0.2cm]
{\color{Brown}\bf\large Abstract}\\
We have investigated the role that different galaxy types have in galaxy-galaxy interactions in compact groups. N-body simulations of 6 galaxies consisting of a differing mixture of galaxy types were run to compare the relative importance of galaxy population demographic on evolution. Three different groups with differing galaxy content were tested: all spiral, a single elliptical and 50\% elliptical. Tidal interaction strength and duration were recorded to asses the importance of an interaction.
A group with an equal number of spiral and elliptical galaxies has some of the longest and strongest interactions with elliptical-elliptical interactions being most significant. These elliptical-elliptical interactions are not dominated by a single large event but consist of multiple interactions. Elliptical galaxies tidally interacting with spiral galaxies, have the next strongest interaction events. 
For the case when a group only has a single elliptical, the largest magnitude tidal interaction is an elliptical on a spiral. Spirals interact with each other through many small interactions.  
For a spiral only group, the interactions are the weakest compared to the other group types. These spiral interactions are not dominated by any singular event that might be expected to lead to a merger but are more of an ongoing harassment.These results suggest that within a compact group, early type galaxies will not form via merger out of an assemblage of spiral galaxies but rather that compact groups, in effect form around an early type galaxy.
\vspace{0.5cm}\\
{\color{Brown}\bf\large Keywords}\\
Galaxy Groups; Galaxy Clusters; Galaxy Evolution; n-body simulations
\vspace{0cm}\\
{\color{Brown}\rule{0.7\textwidth}{2pt}}


\section{Introduction}
Galaxy groups in general, and compact groups in particular are thought to be locations of extreme galaxy evolution due to the very high density of galaxies and therefore lower average spacing between them.  Compact groups typically have $\sim$4 - 6 galaxies with mean separation $\sim$40 kpc and velocity dispersion $\sim$300 km s$^{-1}$. There  is more than circumstantial evidence that interactions take place within compact groups as well as some degree of merger \cite{1-ARAA} and references therein. Correlations between HI content and properties such as galaxy morphology, compactness, and velocity dispersion suggest that  the majority of compact groups are physically dense systems rather than chance projections on the sky \cite{2-AA}. This is consistent with the most compact of groups found to lens background galaxies, indicating the presence of potential well \cite{3-apj}. A linear relation between velocity dispersion and X-ray luminosity for groups and clusters shows that, at the  low end of the X-ray luminosity, inhabited by compact groups, that the correlation has a larger dispersion around the mean and likely contains another component which is non-gravitational (e.g., stellar or AGN sources) resulting from interactions. There are observations that show this to be the case. For example, the intergalactic light (IGL)  surrounding some groups \cite{4-apj} may be a remnant of a high level of star formation in the past based on X-ray studies \cite{5-U}.  Optical images of compact groups show galaxies that are interacting and merging  \cite{6-apj}.  From the 24 velocity fields obtained from a sample of compact groups, rotation curves for 15 galaxies were analyzed.  Only 2 out of the 15 galaxies have rotation curves without any clear evidence of interactions \cite{7-MNRAS}. Ellipticals with highly extended X-ray emission may have formed in groups; the so called "fossil groups". Simulations \cite{8-nat} show that galaxy groups can facilitate the formation of a large elliptical galaxy via mergers, with 80\% of  galaxies in more spherical systems at z  $>$ 0.4 merging into one massive galaxy by z = 0 with a  typical coalescence time-scale of 2-3 Gyr \cite{9-MNRAS}.

While it appears that galaxies interact tidally, the basic nature of the interaction process is still not completely known. \cite{10-PASP} found that 70\% of compact groups are part of larger structures, such as loose groups and galaxy clusters indicating that compact groups are important sites of galaxy evolution in larger structures. Compact groups, whether in larger structures or not, contain a significantly smaller fraction of late-type spiral and irregular galaxies than in their surrounding environment \cite{11-AA}. This strongly supports the conclusion that not only are compact groups a localized site of galaxy evolution, they are likely the site of galaxy mergers. Thus, compact groups likely show various stages of interaction that may precede a merger stage that forms a giant elliptical galaxy.

Cosmological codes are successful in showing that compact groups do form, are numerous and dynamically, relatively long lived. \cite{12-ApJ} found that once galaxies come within 0.5 Mpc of the most massive galaxy, they remain within that distance until z = 0. \cite{9-MNRAS} find a longer typical galaxy merger timescale of 2-3 Gyr, longer than previously believed based on observed abundances of compact groups versus redshift. They conclude that the combination of a large fraction of interlopers and a longer group coalescence timescale alleviates the need for a fast formation process to explain the observed abundance of compact groups (CGs) at z $<$ 0.2. \cite{13-MNRAS} find 20 times as many mock CGs as the HCGs found by Hickson within a distance corresponding to 9000 km/s. This very low (5\%)  Hickson compact group (HCG) completeness is caused by Hickson missing groups that were either faint, near the surface brightness threshold, of small angular size, or with a dominant brightest galaxy. Thus, these simulations indicate that compact groups form in larger numbers and are relatively long lived. Using SDSS and spectral survey data,  \cite{14-ApJ} find that compact groups can be separated into two broad categories: isolated systems and those embedded in rich groups or clusters of galaxies. They find that isolated compact groups have systematically lower dynamical masses than less compact groups at the same group luminosity. This is interpreted as isolated compact groups being at a more evolved stage. On the other hand, compact groups in larger structures may be a mixture of chance alignments in poor clusters and recent infalling groups in rich clusters. Both are observational biases that may account for the difference between CGs in and out of larger structures. This study also suggests that CGs may be the site of galaxy interaction within larger structures in which they occur.


In this paper, we investigate galaxy interactions within compact groups via a simple n-body simulation.  We specifically address the importance of multiple small galaxy interactions versus a few large events as being more influential in a galaxy's evolution.


\section{Compact Group N-Body Interaction Simulation}

We construct a 3-dimensional, n-body simulation of a bound 6 member group of galaxies. In particular, we address the importance of multiple small galaxy interactions versus a single large event as being most influential in a galaxy's evolution via interaction. A second goal is to address this issue in three different morphological settings to determine the relative importance of elliptical versus spiral galaxies as agents of change. The three contrasting morphologies are characterized by the number of elliptical galaxies in the group: no ellipticals, 1 elliptical, and 50\% elliptical. As discussed in the introduction, compact groups typically have elliptical or early type galaxies and are part of larger structures, such as loose groups and clusters of galaxies. The runs consisting of only spirals can be viewed as a pre-compact group with the level of interaction addressing the likelihood of leading to formation of a compact group (i.e., the formation of an elliptical galaxy). Section 2.1 gives the initial conditions and a description of the dynamics of the galaxies. Section 2.2 describes the time step used in the simulation. Section 2.3 describes the tidal interaction between galaxies.

\subsection{Initial Conditions and Dynamics}

The galaxies are initially placed at rest with random coordinates (x,y,z) within a cubic volume of 500 kpc on a side. This value is chosen based on the typical size of CGs when they form in cosmological codes \cite{12-ApJ}. For each
run, only the initial random placement of galaxies changes.
Nominal masses are assigned: 10$^{11}$ Solar masses, typical of a spiral galaxy, and 10$^{12}$ Solar masses, typical
of an elliptical galaxy. While galaxies certainly have a range of masses, we choose to focus on two canonical masses that characterize the two types of galaxies. This approach will best emphasize the roll that mass plays in interaction and dynamics. The motion of a galaxy is determined as follows. Each galaxy has a vector acceleration calculated from the sum of its gravitational
interactions with each of the other galaxies, computed using the Newton's law of gravity. A softening parameter 
is added to the gravitational interaction to account for kinematic energy dissipated through tidal heating that will occur during close interactions. This also reduces the spurious loss of galaxies due to close encounters.
The system evolves using the Newton-Stormer-Verlat leapfrog, or “kick-drift-kick” numerical integration method to advance the particles \cite{15-am}. To determine this force, we use the gravitational equation with the nominal galaxy masses and the separation, R. The separation, R, between bodies in 3 dimensions is given by the root of the sum of the squares of the separation in each dimension with a 4th additional dimension included in this equation, $\epsilon$ \cite{16-ar}, which is the softening parameter. This is shown in equation 1.

\begin{equation}
R^{2} = (\delta x)^{2} + (\delta y)^{2} + (\delta z)^{2}  + \epsilon^{2}
\end{equation}

We use an adaptive epsilon \cite{16-mnras}, which in this simulation varies for each body depending on its mass (M):  $\epsilon$ = M$\times$10$^{-6.5}$. Mass here refers to the mass of a single galaxy. Epsilon is then a separate value for each galaxy, dependent upon its mass. This value is determined in the following way. MASE (Mean Average Square Error) \cite{08-mnras} represents how much the evolution of the system with epsilon differs from the evolution of the system without any epsilon at all. 

\begin{equation}
MASE = \langle\frac{1}{N}\sum_{i=1}^{36}(f_{i}(0) - f_{i}(\epsilon))^{2}\rangle
\end{equation}

Equation 2 describes how MASE is obtained for a given value of epsilon, where f(0) represents the force when epsilon is 0 and f($\epsilon$) represents the force for a given epsilon. For each time step (nominally 1/36 of a run), we take the average square of the difference of the force between every possible interacting pair. The angle brackets show that this quantity is then averaged over a full simulation. This will give a single value for MASE for an entire simulation so that the gravitational interactions are affected by a single value of $\epsilon$. To obtain the data in Figure 1, 9,999 instances of a simulation, all beginning with the same initial conditions, were run simultaneously. Each of these instances of the simulation was governed by a different value of $\epsilon$ and each instance ran its course producing a value for MASE in the end. MASE is plotted against different values of epsilon in Figure 1. We chose the value of epsilon where an increase in epsilon no longer affects the value of MASE. The proportionality constant, 10$^{-6.5}$, is calculated from this value of epsilon and the galaxy mass. To summarize, epsilon reduces the dynamical effect of close galaxy pairs. This would be expected because close galaxies will interact, dissipating kinetic energy. Within the scope of this simulation, the choice of epsilon comes through the calculation of MASE and is an empirical choice, based on its dynamical influence.

\begin{figure}[h]
\includegraphics[width=14.5 cm]{ 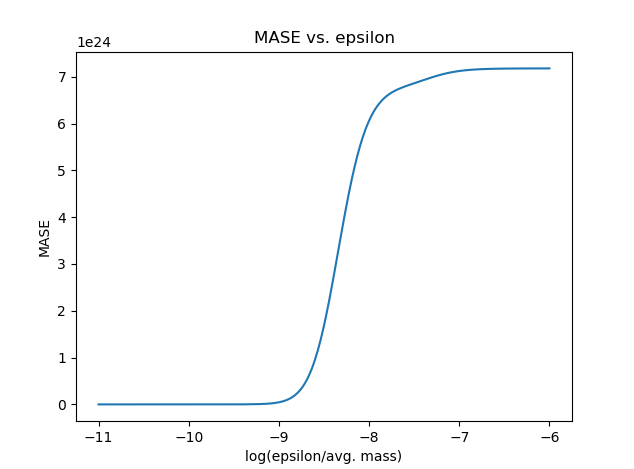 }
\caption{ This shows the change in MASE as epsilon varies. The value of epsilon/mass is the minimum value where it no longer affects the MASE}
\end{figure}  
 
\subsection{Adaptive Time Step}

A simulation runs for 1 Gyr with an initial time step (TS) of 31.71 Myr. This relatively coarse time step, 1/30th of the duration of the simulation, is chosen because at the beginning, the galaxies have an average spacing of 274 kpc and are not strongly interacting. As a result, their mutual dynamical force is not strong and because they begin at rest, initially they do not move very far in a time step. After the initial TS, we use an empirically chosen adaptive TS. In nature, gravity appears to act continually on macroscopic scales so that any timescale used in a simulation is an approximation to gravitational dynamics. The longer the time step the more inaccurate is the approximation to true gravitational dynamics. When galaxies are closer together, the time step is reduced to more accurately simulate the strength and duration of the tidal effect. When the galaxies are further apart and tidal interaction is not important, we lengthen the time step to reduce the overall computational time of the simulation. An additional, and important consideration in using an adaptive time step is that a large time step leads to the loss of galaxies from the simulation volume that is unphysical. That is because, in between time steps, each body experiences a large inertial drift that is unphysical because gravitational acceleration acts continuously. A potential galaxy loss problem is mitigated by introducing an adaptive time step in which the length of the time step is inversely related
to the maximum velocity of the galaxies. This is an intuitive relationship because a higher velocity galaxy will drift more than a slower one and therefore the gravitational acceleration should be
calculated more often to minimize this unphysical drift.  We derive an empirical relationship between time step and maximum galaxy velocity which is given in equation 3. \\

\begin{equation}
TS = \frac{1}{50000\times V{_m}}
\end{equation}

In equation 3, V${_m}$ is the fastest moving galaxy in the simulation in a time step. A scale factor between V$_{m}$ and TS is chosen empirically using many short simulations covering a wide range of scale factors for groups with the three differing galaxy content.  Via this iterative process, the value of 50,000 results in galaxies rarely being ejected from the group. For example, without an adaptive time step, galaxies were frequently lost. Galaxies can still be lost but not due to an artificially long time step. Though this is an empirical treatment, it is adequate for our purposes, which is not a dynamical study of rogue galaxies in groups, but rather a study of  the galaxies in groups. In 30 runs, as we have used in our tidal simulations of each group type, no galaxies are lost. In summary, adapting the time step to avoid artificial rogue galaxies after a close encounter is met using an adaptive TS, which also results in a much more accurate characterization of the tidal interaction.

\subsection{Tidal Interaction}

When a collection of mass components (e.g., gas, stars, and dark matter) forming a self-gravitating ensemble of mass  m${_1}$ passes by another of mass m$_{2}$,
they exert forces on each other. The difference between these forces and the self-gravity holding each galaxy together is the basis of the tidal force. When the mutual force exceeds the self gravitating force, then the galaxies
act to pull each other apart. Equation 4 shows the inequality that must be met for a tidal interaction \cite{duc-19}. The strength of the tidal interaction is the magnitude of the difference in the  terms  on the right and left of the inequality.

\begin{equation}
\frac{Gm_1}{r_2^2} <  Gm_2(\frac{1}{(r_1 - r_2)^2} - \frac{1}{{r_1}^2}) 
\end{equation}

The galaxies are approximated as falling toward each other.  In equation 4, r${_1}$ is the center-to-center distance between the two interacting galaxies.  The radius of the galaxy that is being tidally affected is r${_2}$.  The values for radius and mass are given nominal values for spiral and elliptical galaxies. Spiral galaxies are given a radius of 25 kpc, and mass 10$^{11}$ Solar masses and elliptical galaxies are given radius 50 kpc and mass 10$^{12}$ Solar masses. While both types of galaxies take on a range of masses in nature, assigning these physically meaningful canonical values facilitates simulating the dependence  of tidal interaction on galaxy mass. The strength is recorded for each time step when the inequality is met. The time step is also recorded. Since we know the mass for each galaxy type, the tidal effect is recorded as an acceleration. We show the cumulative tidal acceleration between galaxy types and the duration of the  level of  tidal acceleration in the figures in Section 3. The lateral tidal component, not modeled here, also occurs when this inequality is met, and may be important in increasing the rate of star formation in the galaxies \cite{96-hen}. Thus, tidal interaction may be an important evolutionary component of galaxies in close proximity.

\section{Results}%
\subsection{Distribution of Tidal Interactions}

When there are only spiral galaxies in the group, the distribution of tidal interaction strength is symmetric around the mean because all of the spirals have the same mass (Figure 2).

\begin{figure}[h]
\centering
\includegraphics[width=14.5 cm]{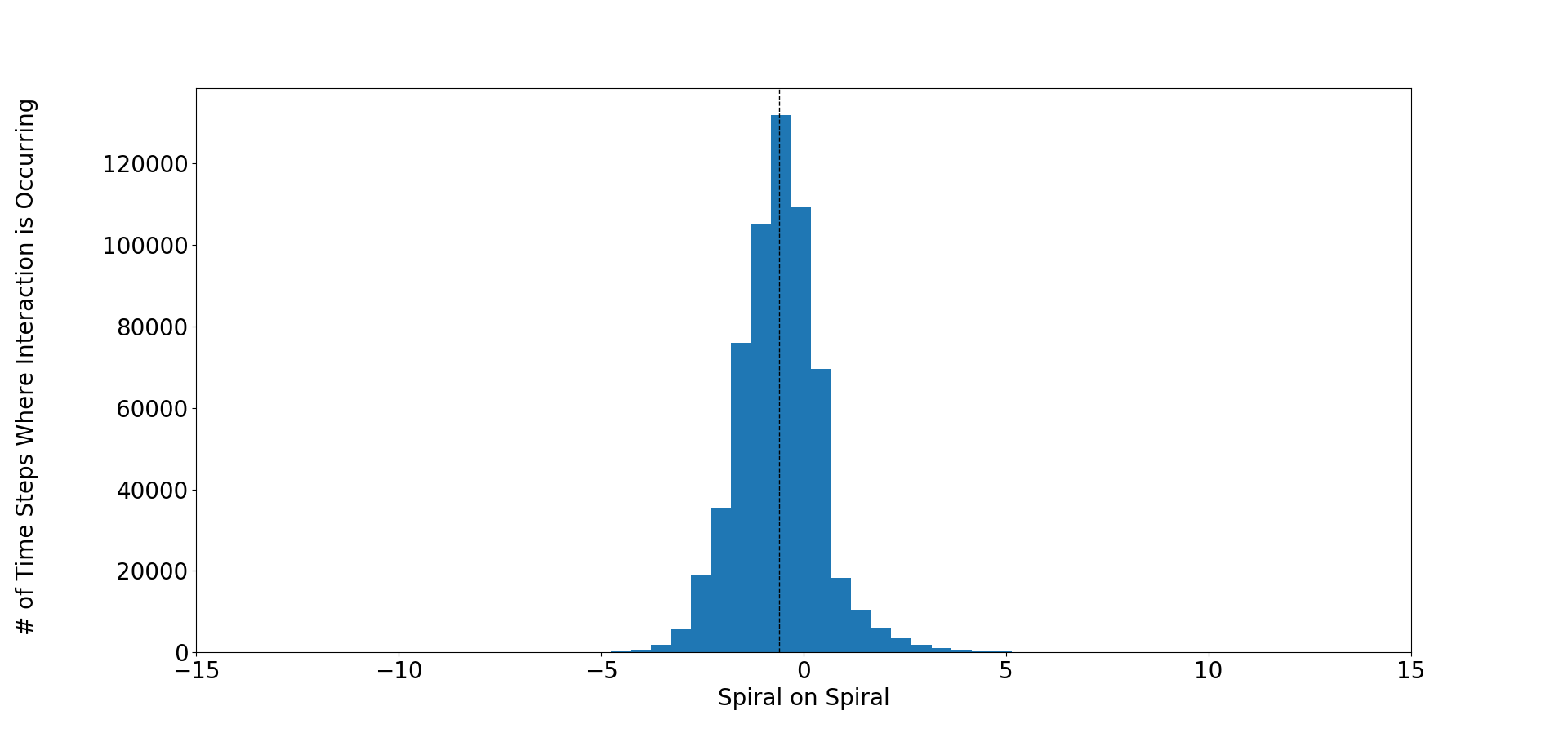}
\caption{Distribution of tidal interaction strength (Mpc Ps$^{2}$), where Ps is 10$^{15}$ seconds, between galaxies in a group composed only of masses typical of a spiral galaxies. The mean is near the peak of a symmetric distribution indicating that a few strong interactions
do not dominate the interactions.}
\end{figure}  
\begin{figure}[ht]
\begin{minipage}[b]{.5\textwidth}
\centering
\includegraphics[width=1.1\textwidth, height=1.25\textwidth]{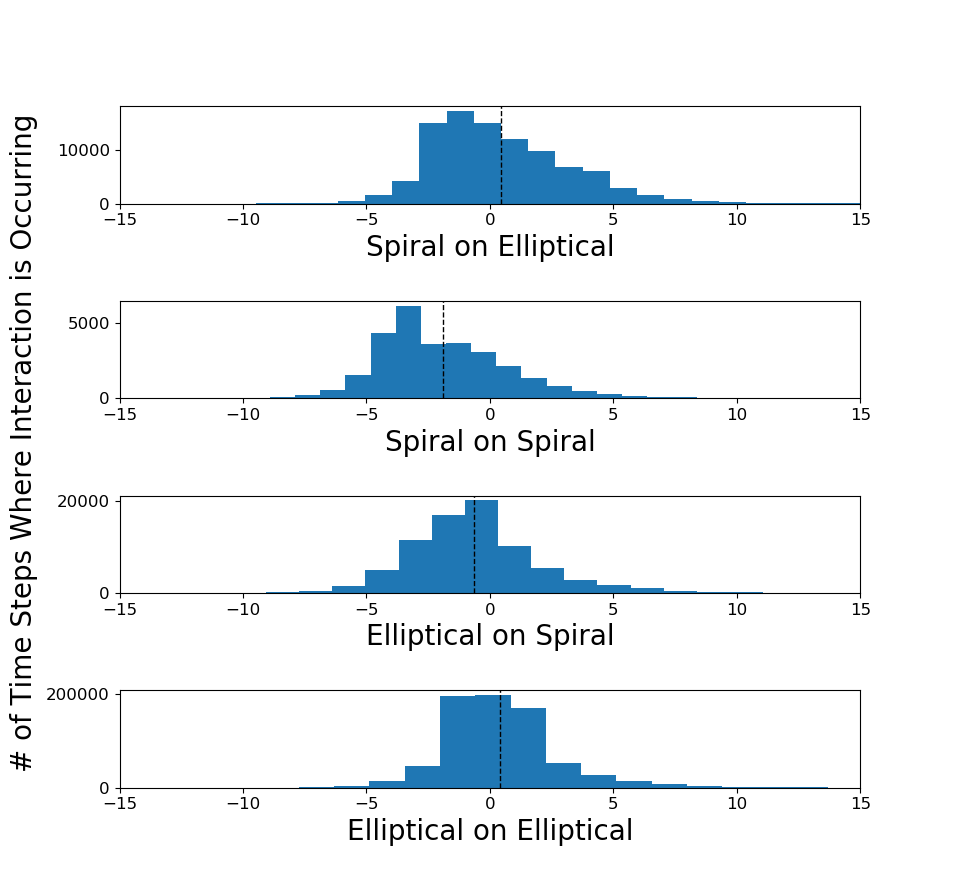}
\caption{Distribution of tidal interaction strength between galaxies for a group with 50\% masses typical of spiral and 50\% typical of ellipticals. The mean is shown with a vertical dotted line}
\end{minipage}
\hfill
\begin{minipage}[b]{.5\textwidth}
\centering
\includegraphics[width=1.1\textwidth, height=1.25\textwidth]{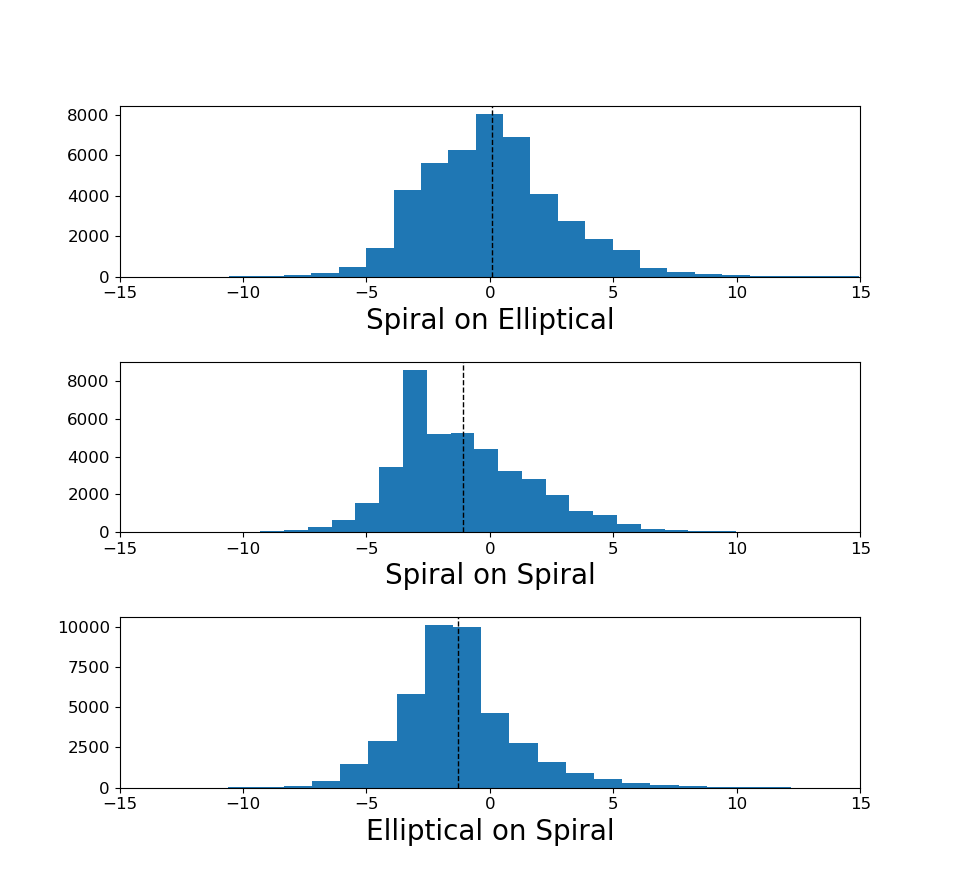}
\caption{Distribution of tidal interaction strength between galaxies for a group with 5 masses typical of spiral galaxies and 1 typical of an elliptical. The mean is shown with a vertical dotted line}
\end{minipage}
\end{figure}

Figures 3 and 4 show 3 elliptical galaxies and 1 elliptical galaxy in the group, respectively. This data is for a single run. The distribution of interaction strengths and the mean (shown by the dotted line) is plotted for each combination of galaxy type.
The mean is shifted away from the peak of the distribution for interactions in a group with at least one elliptical galaxy. This is in contrast to the spiral only group and indicates that the presence of an elliptical galaxy changes the
spectrum of galaxy interactions. This is expected due to the larger mass of an elliptical in a mixed pair. Much earlier, \cite{hen-99} found that for a given L$_{B}$, early-type galaxy pairs were underluminous in diffuse X-ray emission compared to isolated galaxies. This was interpreted as resulting from interaction.
What is most striking in these simulations is the large offset between the distribution and
the mean strength for elliptical-elliptical interactions in Figures 3 and 4. This is also seen to a lessr degree in the other interactions involving an elliptical. The offset is due to a few strong events dominating the interactions involving elliptical galaxies.
The strength at which the peak number of interactions occurs varies with pair type. For example, the peak number for spiral-elliptical interactions is farther to the left of the mean (i.e., weaker) for 50\% elliptical  groups than when there is a single elliptical in the group. This means that spirals interact with ellipticals more weakly when there are more ellipticals in the group. The interpretation is that the mean distance between a spiral and  elliptical is larger in a 50\%  elliptical group. For spiral-spiral interactions, the peak number is farther to the right of the mean (stronger) for 50\% ellipticals. This means that when there are more ellipticals in the group, spirals interact with spirals more strongly than when there are fewer ellipticals. The interpretation is that the mean separation between spirals is smaller in elliptical dominated groups. To summarize, in a 50\% elliptical group, spiral pairs tend to be closer together than spiral-elliptical pairs. This may be due to
the dynamical effect of the higher mass ellipticals having a slingshot effect on spirals. Thus, dynamics, which result from galaxy content in the group, effect the interactions.


\subsection{Tidal Interaction Strength in Multiple Runs}

\begin{figure}[h]
\includegraphics[width=14.5 cm]{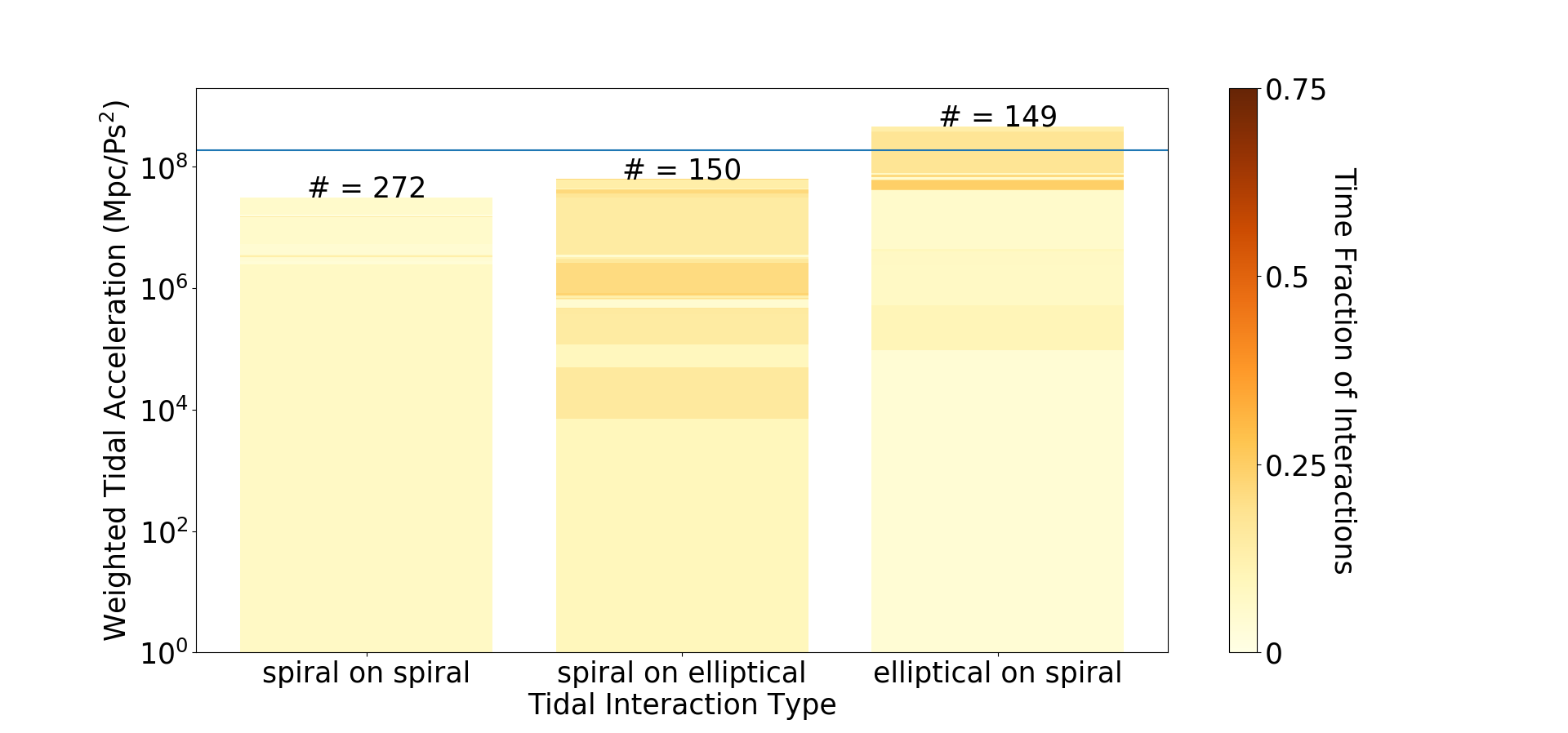}
\caption{This group is spiral dominated with a single elliptical galaxy. Tidal acceleration for each run is  color coded for length of
   interaction. The elliptical on spiral has few interactions but are some of the longest interactions. In contrast, the spiral on spiral are weaker, of shorter duration but much more numerous.}
\end{figure}  
We expand on the concept of galaxy harassment in structures \cite{moo-96} whereby galaxies experience numerous encounters as part of a dynamically evolving structure by addressing the dependence of interaction on galaxy mass.  When a compact group consists of 5 spirals and a single elliptical, the largest magnitude tidal interactions are dominated by those involving an elliptical on a spiral (Figure 5). In this figure, the strength of the tidal interaction is cumulative and shown on the
verticle axis. Each tidal acceleration (in units of Megaparsec per petasecond) is given a color in which the color is indicative of the duration of interaction. The duration is a sum of the time steps for a given interaction strength. Yellow is the shortest duration and orange longer. We also show the number of interactions which allows evaluating whether the interactions between galaxy type are dominated by a few strong events (we call bullying) or many weaker interactions (we term harassment). We find that  in 8 out of 30 runs, the largest tidal interaction in a run is a spiral on an elliptical while in 20 out of 30 runs the largest tidal interaction is an elliptical on a spiral. In this type of group, we view ellipticals as being bullies because they have larger overall impact (the product of strength of interaction and duration) with minimum interaction number. At the other extreme (least overall impact with most encounters), spirals merely harass each other, as they pose much less of a physical threat to each other compared to the elliptical. An elliptical bullying a spiral may lead to morphological changes inferred from the mismatch in mass,  but a merger may be more likely between spirals as they spend more time together, a dynamical condition that favors merger. For a spiral interacting with the elliptical, the tidal interaction is less extreme due to the mismatch in mass. In this case, the spiral galaxies take on the role of harassers. 
\begin{figure}[h]
\includegraphics[width=14.5 cm]{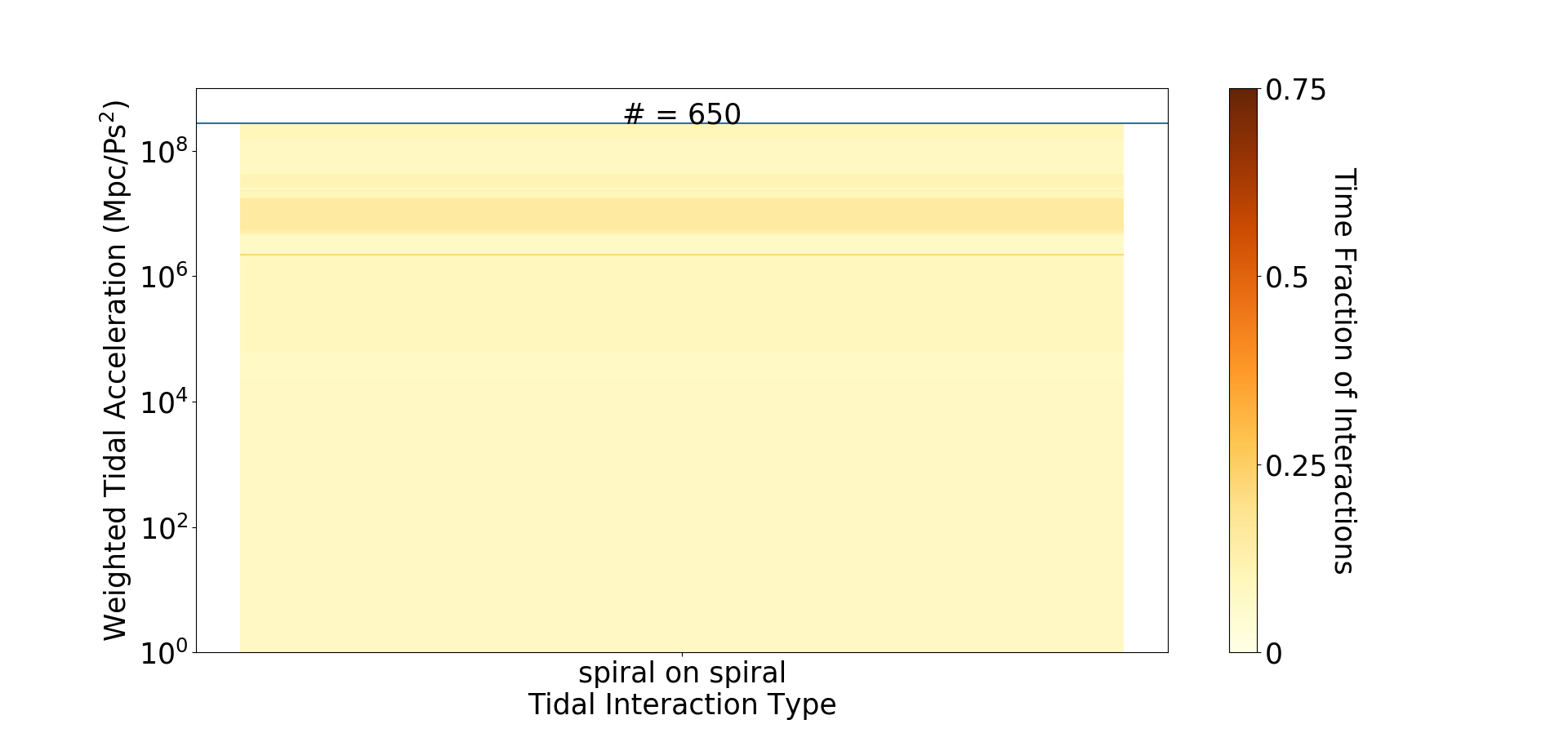}
\caption{This group is made up entirely of spiral galaxies. Shown is the tidal acceleration for each run, color coded for length of interaction.}
\end{figure}  
Figure 6 shows interaction strength, based on tidal acceleration between galaxies,
for 30 iterations of a spiral only galaxy group. From comparison with Figure 5, it is apparent that spiral-spiral interactions, are generally the weakest interactions and of short duration. It is unlikely that spiral-spiral interactions lead to mergers in a spiral only group. This is due to the overall lower level of interaction resulting from the dynamics of a spiral only group. Removing the largest interaction from each run, we see that there is no noticeable change in magnitude of the interactions suggesting that the spiral on spiral interactions are not dominated by any strong, singular event, and that dynamically, mergers are unlikely. As this is different than for a group with a single elliptical, we conclude that the diminished interaction is due to the difference in dynamics due to the presence of an elliptical and that spiral-spiral mergers are more favorable in groups with at least one massive galaxy.
\begin{figure}[h]
\includegraphics[width=14.5 cm]{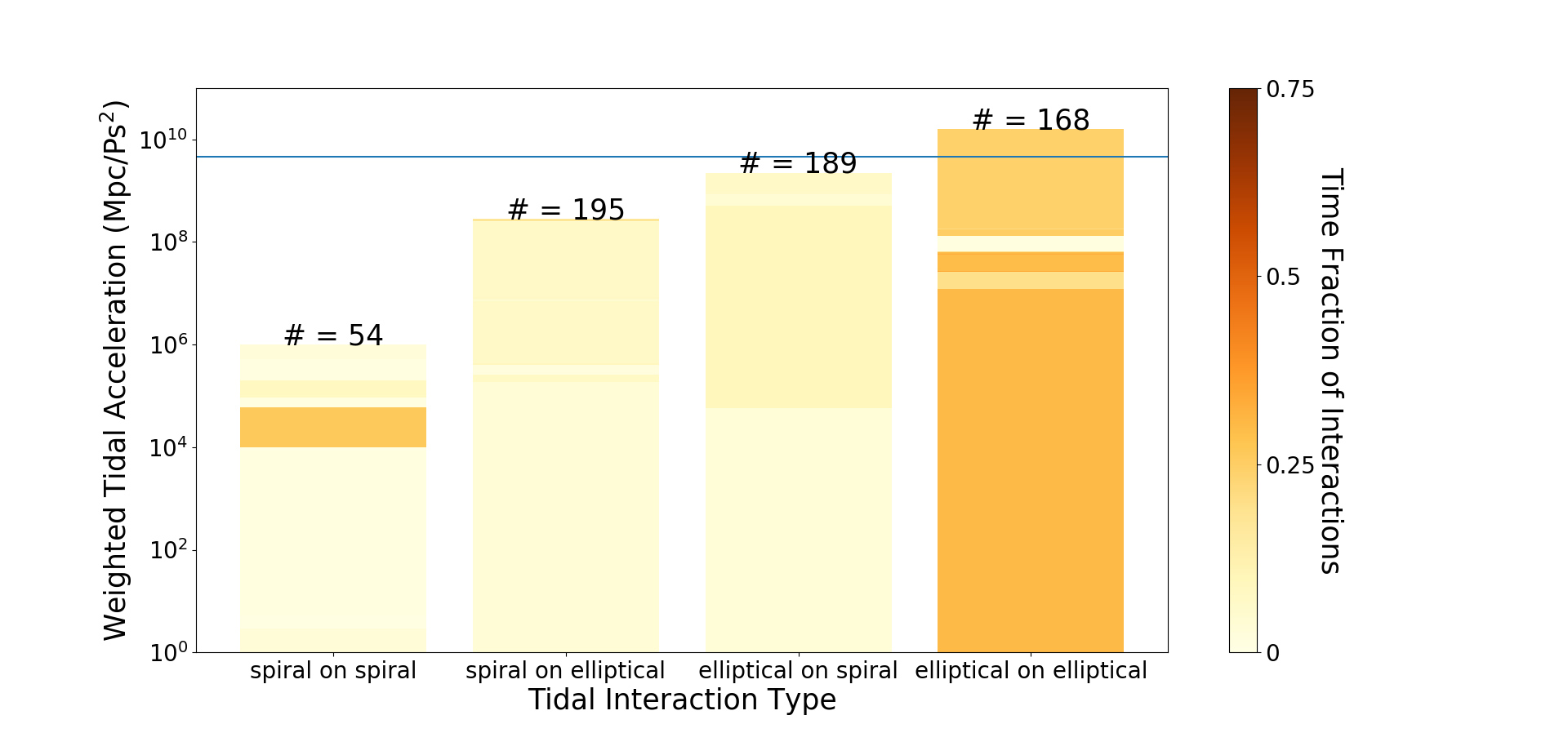}
\caption{This group is made up 50\% spiral galaxies and 50\% elliptical galaxies. This shows the tidal acceleration for each run, color coded for length of interaction. Interactions between Elliptical galaxies are the longest and the strongest.}
 \end{figure}  
Figure 7 shows interactions for a group with 50\% spirals and 50\% ellipticals. This group has the longest and strongest interactions. Within this morphological type, elliptical-elliptical interactions are the strongest and longest. It is difficult to predict the consequences of such strong and long interactions though it is likely to lead to significant evolution, perhaps through morphological changes in the ellipticals. A distant second in interaction strength is ellipticals on spirals. We also note that the elliptical-elliptical interactions during a run are not dominated by a single large event. This is verified from over 30 runs. Interestingly, in 20\% of the runs, the strongest tidal interaction in a run, is a spiral harassing an elliptical. This implies
that, though rare, in some groups a significant fraction of ellipticals might experience significant evolution from tidal encounters with spirals.
\eject
\section{Discussion}
Interaction between galaxies is believed to be linked to significant changes in the galaxy population. Interaction appears to be a widespread phenomenon in the Universe because at low redshift we see bound companions \cite{18-ARAA} and interacting galaxies. Properties such as galaxy metallicity are linked to environmental factors such as high galaxy density. For example, higher mass galaxies are observed to be less affected by their environment compared to low mass galaxies, based on studies of metallicity gradient \cite{19-mnras}. Both ram-pressure by an intergalactic medium and tidal interactions pit the environment against the galaxy's own restoring, gravitational potential. A higher mass galaxy is nominally more likely to resist environmental forces in both mechanisms. While this difference in metallicity with mass is consistent with the presence of an environmental effect that is more successful with a lower mass galaxy, it doesn’t favor or disfavor ram-pressure verses galaxy interaction. However, there is substantial support for galaxy evolution due to strangulation \cite{20-nat} from gas loss associated with tidal effects by the cluster potential. Galaxies in compact groups also appear to be deficient in neutral hydrogen compared to isolated galaxies of similar properties which is indicative of an environmental stressor. There is sufficient evidence that ram-pressure as well as galaxy interactions are important processes in both clusters and groups \cite{14-gavv}. However, processes such as ram pressure and strangulation are insufficient to explain this observed HI deficiency \cite{20-mnras}. Interactions may be isolated, repeated, or the prelude to a catastrophic merger that has a deep impact on the galaxy stellar mass function \cite{14-mnras}. The morphological transformation of a galaxy from mergers is dependent on minor mergers, more so for ellipticals. \cite{18-mnras} find that elliptical galaxies have undergone more mergers than disc-like galaxies and that the highest-mass galaxies are dominated by merging with lower-mass galaxies. The brightest group galaxies are found through observation to grow in stellar mass due to mergers but at a lower rate  than the stellar mass growth predicted by galaxy evolution models \cite{21-apj}. \cite{23-mnras} characterize the interaction stage frequency between galaxies as the following. Close pair fraction is 9\% and is identified with early stage. Close pairs with tidal feature is 5\% and is an intermediate stage. The late stage is a tidal feature, 23\%. This significant fraction of galaxies in close proximity and tidally interacting indicates that tidal interaction plays an important role in galaxy evolution and perhaps merger dynamics. 

We \cite{uni-21} previously investigated the origin of intergalactic light (IGL) in nearby compact groups of galaxies. IGL is hypothesized to be the byproduct of interaction and merger within compact groups. The X-ray point source population in a sample of compact groups that have IGL were compared with compact groups without IGL. The disk sources were found to have X-ray luminosities typical of high mass X-ray binaries (HMXRBs). A comparison of the luminosity distribution of HMXRBs with compact groups point sources shows a shift toward lower luminosity in compact groups.  Since HMXRB luminosity is proportional to the star formation rate (SFR) we inferred that the galaxies in groups with IGL show a trend toward reduced SFR. We interpret this to mean that groups with visible IGL represent a later evolutionary phase than compact groups without IGL. This is due to their galaxies experiencing a quenching of star formation, inferred from lower HMXRB luminosities. Simulations also show that the presence of IGL is post-starburst, a phase that accompanies the gas removal mechanism that creates the IGL. In fact, intergalactic light has lead to the discovery of a group of tidally interacting galaxies \cite{gira-23}, emphasizing that compact groups are important sites of galaxy interaction.

We have simulated and quantified the tidal interaction between galaxies in a compact group to understand how it depends on varying galaxy content. In particular we address the relative importance of many small interactions versus a few very large ones. A group consisting of equal number of spirals and ellipticals has the longest and strongest interactions. Within this morphological type, elliptical-elliptical interactions are characterized by long duration, strong tidal encounters.  In any single run, a few singularly strong events are seen to dominate the interactions. Such long and strong interactions between ellipticals is likely to lead to significant evolutionary and morphological changes in the ellipticals. A distant second in interaction strength is ellipticals on spirals. Conversely, spirals have a weaker accumulated interaction on ellipticals, though in 20\% of the runs, the strongest tidal interaction in the run is a spiral harassing an elliptical. There are more interactions between spirals than observed in either ellipticals or mixed pairs.  The interpretation is that the mean distance between a spiral and  an elliptical is larger in the 50\%  elliptical group and the mean separation between spirals is smaller. Spirals tend to be closer together than spirals and ellipticals due to the dynamical effect of the higher mass ellipticals having a slingshot effect on spirals so that they spend more time, on average, segregated by galaxy type. 
When a group only has a single elliptical, the largest magnitude tidal interactions are those involving an elliptical on a spiral. In 20 out of 30 runs the largest tidal interaction is an elliptical on a spiral compared to a mere 8 out of 30 runs, in which the largest tidal interaction is a spiral on an elliptical. In these elliptical poor groups, we might view an elliptical as a bully, while spirals merely harass each other over time. While an elliptical bullying a spiral may lead to morphological changes, a merger may be more likely between spirals as they spend more time together. For a spiral interacting with the elliptical, the tidal interaction is less extreme due to the mismatch in mass. In this case, the spiral galaxies take on the role of minor harassers. Spiral-spiral interactions in a spiral only group are the weakest interactions in any of the 3 group morphologies simulated. It is unlikely that spiral-spiral interactions lead to mergers in a spiral only group. This is because the strongest spiral on spiral interactions are not dominated by any singular event that might be expect to lead to a merger. The overall, lower level of interaction compared to group with a single elliptical suggests that the diminished interaction is due to the difference in dynamics. These results are consistent with observational studies. Galaxy interactions in clusters of galaxies show that interactions with a nearby, early-type galaxy is the main driver in quenching star formation in late-type spirals, rather than the larger cluster environment \cite{park-2}. A large sample of galaxy pairs in the CFA survey shows that the luminosity of the companion galaxy is the crucial parameter in determining interaction effects due to tidal forces \cite{wood-1}. As luminosity scales with mass in galaxies, this is consistent with our result that an elliptical galaxy in a simulated compact group is the primary agent of interaction.

Galaxies are not randomly distributed in the universe but are clumped into groups and clusters \cite{AA-17}. The  substructure within clusters is most easily seen in X-ray maps that commonly exhibit discreet substructures as well as an asymmetric  shape that reflects the underlying galaxy substructure. Even within clusters of galaxies, galaxies are not randomly distributed and are clumped into substructures ranging from pairs and groups up to subclusters. This trend extends to the scale of superclusters in which the X-ray emission is identified with clusters, groups, and filaments \cite{tit-01}. The richest catalogued detail of supercluster content is the Laniakea Supercluster, which is characterized by clusters and many galaxy groups \cite{nat-14}. On all scales, galaxies interact with other galaxies in close proximity within groups so that simulations of galaxy evolution within compact groups, as we have done here, is likely applicable to the evolution of galaxies during cluster formation via the cosmic web.

\def\cprime{$'$} \def\cprime{$'$}
{\color{Brown}
\end{document}